%% file: main.tex
\documentclass[xcolor=pdftex,usenames,dvipsnames,table]{PoS}

\usepackage{slashed}
\usepackage{graphicx}
\usepackage{amssymb,amsmath}
\usepackage{pifont}
\usepackage{multirow,lscape}
\usepackage{array}
\usepackage[lofdepth,lotdepth]{subfig}
\usepackage{verbatim}
\usepackage{bm}
\usepackage{comment}

\graphicspath{{./figs/}}

\providecommand{\abs}[1]{\lvert#1\rvert}
\providecommand{\expv}[1]{\langle#1\rangle}
\providecommand{\bv}[1]{\boldsymbol{#1}}
\providecommand{\wbar}[1]{\overline#1}
\newcolumntype{C}[1]{>{\centering\arraybackslash}p{#1}}

\definecolor{indigo}{RGB}{75,0,130}
\definecolor{darkorange}{RGB}{255,140,0}
\definecolor{darkgreen}{RGB}{0,102,68}

\definecolor{gray}{rgb}{0.8,0.8,0.8}
\newcommand{\snu}[1]{} 

\title{%
Heavy-Meson Spectrum Tests of the Oktay--Kronfeld Action}

\ShortTitle{%
Heavy-Meson Spectrum Tests of the OK Action}

\author{%
  Jon A. Bailey, \speaker{Yong-Chull Jang}, Weonjong Lee \\
  Lattice Gauge Theory Research Center, CTP, and FPRD, \\
  Department of Physics and Astronomy, 
  Seoul National University,
  Seoul, 151-747, South Korea\\
  E-mail: \email{wlee@snu.ac.kr}
}

\author{%
  Carleton DeTar\\
  Department of Physics and Astronomy, University of Utah,
  Salt Lake City, UT  84112, USA\\
  E-mail: \email{detar@physics.utah.edu}
}

\author{%
  Andreas S. Kronfeld\\
  Theoretical Physics Department, Fermilab, Batavia, IL  60510, USA\\
  Institute for Advanced Study, Technische Universit\"at M\"unchen,
  85748 Garching, Germany\\
  E-mail: \email{ask@fnal.gov}
}

\author{%
  Mehmet B. Oktay\\
  Department of Physics and Astronomy, University of Iowa, Iowa City, IA 52242, USA 
}

\author{Fermilab Lattice, MILC, and SWME Collaborations}

\abstract{ 
We present heavy-meson spectrum results obtained using the Oktay--Kronfeld
(OK) action on MILC asqtad lattices.
The OK action was designed to improve the heavy-quark action of the Fermilab
formulation, such that heavy-quark discretization errors are reduced.
The OK action includes dimension-6 and -7 operators necessary for
tree-level matching to QCD through order $\mathrm{O}(\Lambda^3/m_Q^3)$ for
heavy-light mesons and $\mathrm{O}(v^6)$ for quarkonium, or, equivalently,
through $\mathrm{O}(a^2)$ with some $\mathrm{O}(a^3)$ terms with Symanzik
power counting.
To assess the improvement, we extend previous numerical tests with
heavy-meson masses by analyzing data generated on a finer ($a \approx
0.12\;$fm) lattice with the correct tadpole factors for the $c_5$ term
in the action.
We update the analyses of the inconsistency parameter and the hyperfine
splittings for the rest and kinetic masses.
}

\FullConference{%
  The 32nd International Symposium on Lattice Field Theory\\
  23-28 June, 2014\\
  Columbia University, New York, NY}

\begin{document}

\section{Introduction}

The parameter $\epsilon_K$ quantifies indirect CP violation in the
neutral kaon system.
%
%
%
At present, the tension between the Standard Model (SM) and
experimental values of $|\epsilon_{K}|$ is $3.4\sigma$~\cite{Wlee} with the value of
$|V_{cb}|$ from the exclusive decay $B\to D^*\ell\nu$~\cite{JWL}.
This value of $|V_{cb}|$, the most precise from exclusive decays to date, is
$3\sigma$ away from the value from inclusive decays~\cite{GS}.
The largest error in the $\epsilon_K$ determination in the SM comes
from $|V_{cb}|$, so it is crucial to improve the precision of
exclusive~

The dominant error of exclusive $|V_{cb}|$ comes from the heavy-quark
discretization error in the form-factor calculation of the
semi-leptonic decay $B\to D^*\ell\nu$ \cite{JWL}. 
Hence, the SWME Collaboration plans to use the Oktay--Kronfeld (OK) action~\cite{OK} in the
upcoming calculation in order to reduce it efficiently.
This action is an improved version of the Fermilab action~\cite{EKM},
which incorporates the dimension-6 and -7 bilinear operators needed
for tree-level matching to QCD through order $\mathrm{O}(\Lambda^3/m_Q^3)$
for heavy-light mesons and $\mathrm{O}(v^6)$ for quarkonium.
We expect that the bottom- and charm-quark discretization errors could
be reduced below the current $1\%$ level.
A~similar error for the
charm-quark could also be achieved with other highly-improved
actions, such as HISQ~\cite{Follana:2006rc}.

For the heavy-meson spectrum, we present results for the inconsistency
parameter \cite{Collins,Kronfeld} and hyperfine splittings, all of which
test how well the Fermilab and OK actions perform in practice.
For this purpose, we follow the strategy of our previous
work~\cite{MBO:LAT2010}, in which the $c_5$ term was not
completely tadpole-improved.
In this work, we now implement the tadpole improvement for $c_5$
completely.
We also extend the data analysis to data sets produced on a finer ($a \approx
0.12\;$fm) MILC asqtad lattice.

\section{Meson Correlators}

\input{tables/ensemble}

We use a subset of the MILC $N_f=2+1$ asqtad ensembles at $a=0.12\;$fm and
$0.15\;$fm~\cite{Bazavov:RevModPhys.82.1349}, summarized in
Table~\ref{tbl:ensemble}.
We compute meson correlators $C(t,\bm{p})$
\begin{equation}
  \label{eq:corr}
	C(t,\bm{p}) 
	  = \sum_{\bm{x}} e^{\mathrm{i}\bm{p} \cdot \bm{x}}
		\expv{\mathcal{O}^{\dagger}(t,\bm{x}) \mathcal{O}(0,\bm{0})} \,.
\end{equation}
The interpolating operators $\mathcal{O}(x)$ are
\begin{align}
  \mathcal{O}_\mathsf{t}(x) 
  &= \bar{\psi}_\alpha(x) \Gamma_{\alpha\beta} 
  \Omega_{\beta\mathsf{t}}(x) \chi(x)
  & \text{(heavy-light meson)} \,,&
  \\
  \mathcal{O}(x) 
  &= \bar{\psi}_\alpha(x) \Gamma_{\alpha\beta} 
  \psi_{\beta}(x) & \text{(quarkonium)}\,, & 
\end{align}
where the heavy-quark field $\psi$ is that of the OK action, while
the light-quark field $\chi$ is that of the asqtad action.
The spin structure is $\Gamma = \gamma_5$ for the pseudoscalar and
$\Gamma = \gamma_i$ for the vector meson.
The taste degree of freedom for the staggered fermion is obtained from
the 1-component field $\chi$ with $\Omega(x) = \gamma_1^{x_1}
\gamma_2^{x_2} \gamma_3^{x_3} \gamma_4^{x_4}$ \cite{Wingate,Bernard}.

We compute 2-point correlators with 4 different values of hopping
parameter: $\kappa =$ 0.038, 0.039, 0.040, 0.041.
We fix the valence light-quark mass to $am_s$ in Table~\ref{tbl:ensemble}.
We choose 11 meson momenta, $a\bv{p} = (2\pi/N_L) \bv{n}$:
$\bm{n} =$ (0,0,0), (1,0,0), (1,1,0), (1,1,1), (2,0,0), (2,1,0), 
(2,1,1), (2,2,0), (2,1,1), (3,0,0), (4,0,0).
To increase statistics, correlators are computed at 6 different
source time slices on each gauge configuration.

\begin{figure}[t]
  \centering
  \subfloat[Residual]{%
 	  \includegraphics[width=.45\textwidth]
                          {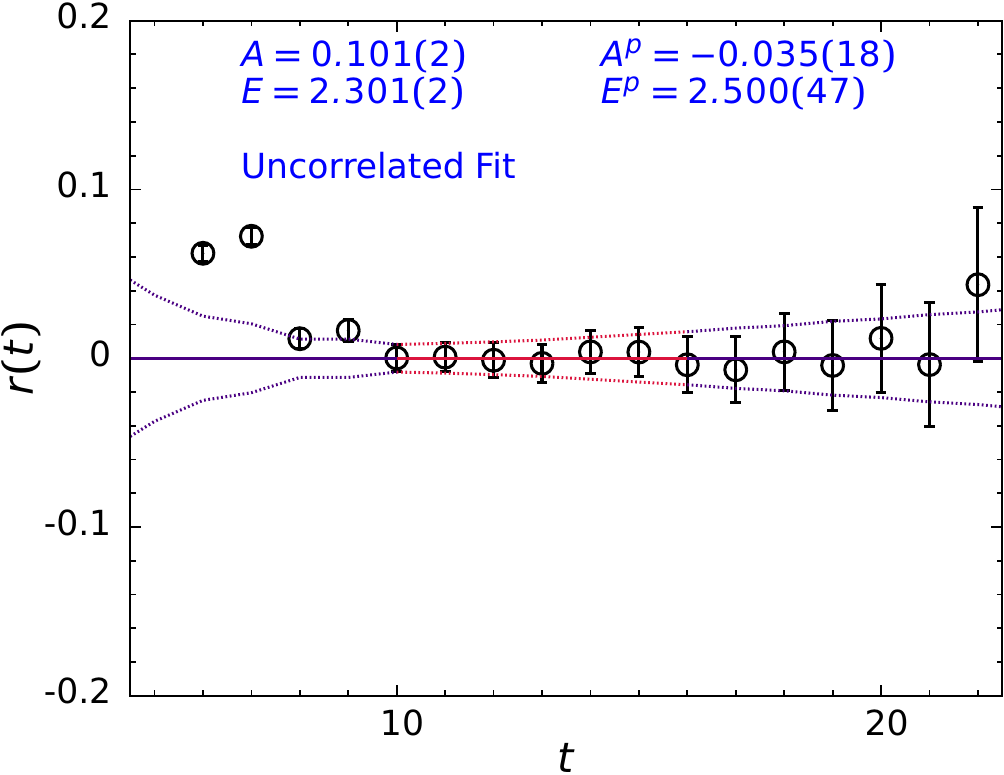}
  }
  \hfill
  \subfloat[Effective Mass]{%
 	 \includegraphics[width=.45\textwidth]
                         {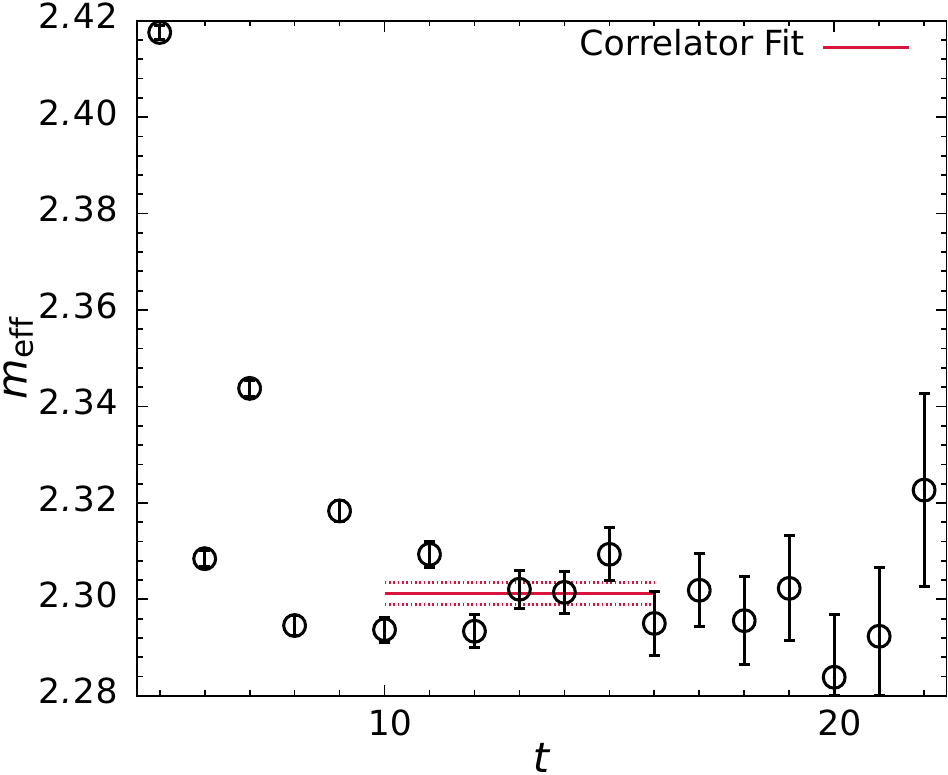}
  }
  \caption{ $r(t)$ and $m_\text{eff}(t)$ for a pseudoscalar
    heavy-light meson correlator at $\kappa=0.038$ and
    $\bv{p}=\bv{0}$, obtained using the uncorrelated fit.  }
  \label{fig:corr-fit}
\end{figure}

Each correlator is folded in half, and then fit to the function
\begin{equation}
  f(t) = A \{e^{-E t} + e^{-E (T-t)}\} 
       +(-1)^t A^p \{e^{-E^p t} + e^{-E^p (T-t)}\} \,,
\end{equation}
where $A$, $A^p$, $E$, and $E^p$ are determined by \snu{the} fitting.
Figure \ref{fig:corr-fit} shows the correlator fit results with fit
residual $r(t)$ and effective mass $m_\text{eff}(t)$ for a
pseudoscalar heavy-light meson data:
\begin{equation}
	r(t) = \frac{C(t)-f(t)}{\abs{C(t)}} \,,\qquad
	m_\text{eff}(t) = \frac{1}{2} \ln \Bigg(\frac{C(t)}{C(t+2)}\Bigg)
        \,.
\end{equation}
We exclude the largest momentum, $\bm{n}=(4,0,0)$, from the dispersion
relation fits, because these data are very noisy, and the correlator
fits are poor.


\section{Dispersion Relation}
\label{sec:disp-rel}

Once we obtain the ground state energy at each momentum, we fit them
to the non-relativistic dispersion relation \cite{EKM},
\begin{equation}
  \label{eq:disp}
  E = M_1 + \frac{\bm{p}^2}{2M_2} - \frac{(\bm{p}^2)^2}{8M_4^3}
     - \frac{a^3 W_4}{6} \sum_i p_i^4 \,,
\end{equation}
where $M_1$ ($M_2$) is the rest (kinetic) mass.
In the Fermilab formulation, the kinetic mass is chosen to be the
physically relevant mass \cite{EKM}, because that choice minimizes discretization errors in matrix elements 
and in mass splittings.

\begin{figure}[t]
  \centering
  \subfloat[Heavy-light]{%
    \includegraphics[width=.45\textwidth]
                    {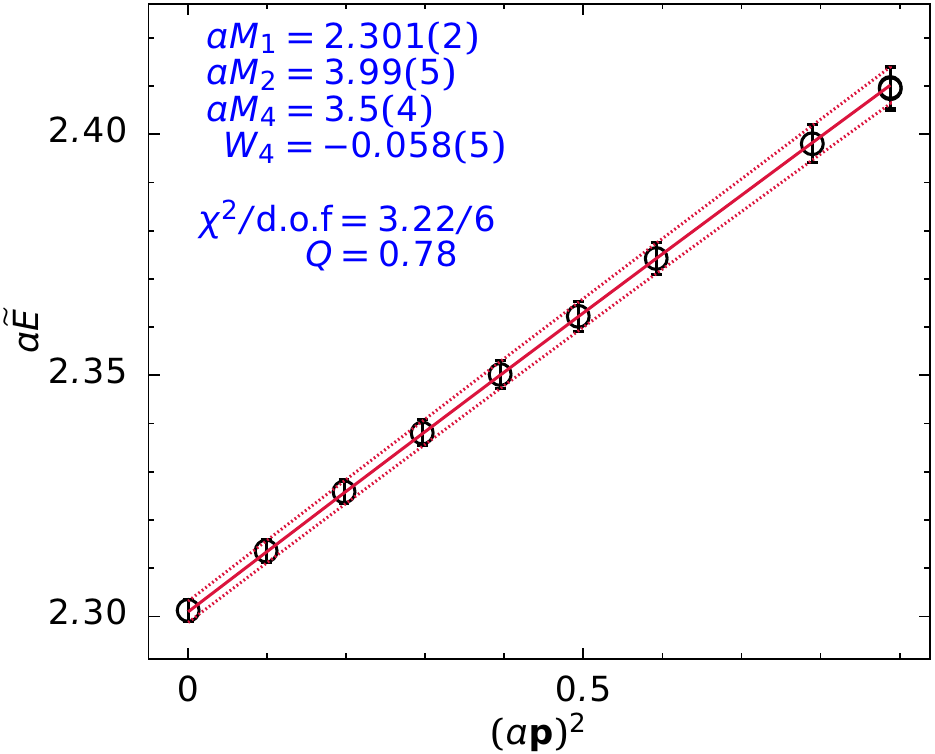}
  }
  \hfill
  \subfloat[Quarkonium]{%
    \includegraphics[width=.45\textwidth]
                    {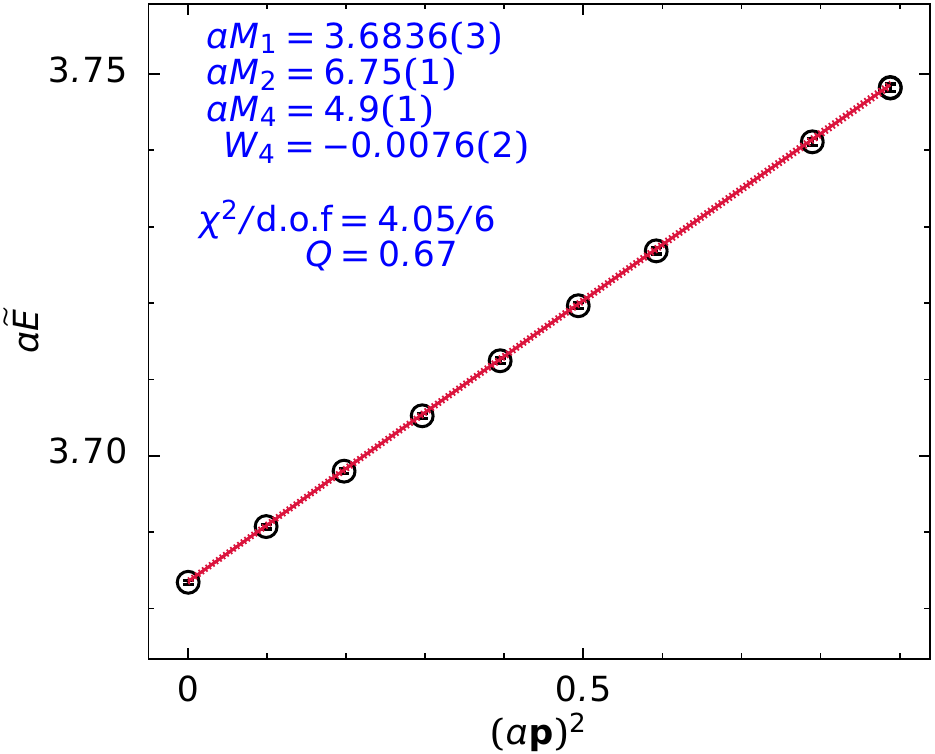}
  }
  \caption{ Fit results of pseudoscalar meson spectrum to dispersion
    relation at $\kappa=0.038$. }
  \label{fig:disp-ps}
\end{figure}

When fitting the data to the dispersion relation, we use the full
covariance matrix and no Bayesian prior information.
In Fig.~\ref{fig:disp-ps}, we plot results after subtracting from the
data the $W_4$ term, which parametrizes the breaking of O(3) rotational symmetry.
Here, $\widetilde{E}$ is defined to be
\begin{equation}
  \label{eq:disp-noart}
  \widetilde{E} = E + \frac{a^3 W_4}{6} \sum_i p_i^4 \,.
\end{equation}
Note that the two data points at momenta $\bv{n} = (2,2,1)$ and
$(3,0,0)$ lie on top of each other due to the removal of the $W_4$
term.
As one can see from the plots, fits to Eq.~\eqref{eq:disp} are good enough to
determine the kinetic mass reliably.

\section{Inconsistency Parameter}

The inconsistency parameter $I$, Eq.~\eqref{eq:iparam}, is designed to
examine the improvements by $\mathrm{O}(\bm{p}^4)$ terms in the
action~\cite{Collins,Kronfeld}.
This is, in particular, good for probing the improvement by the OK action,
because it isolates those improvement terms by construction.
\begin{equation}
  \label{eq:iparam}
  I \equiv \frac{2{\delta}M_{\wbar{Q}q} 
                 - ({\delta}M_{\wbar{Q}Q} + {\delta}M_{\wbar{q}q})}
				{2M_{2\wbar{Q}q}}
	= \frac{2{\delta}B_{\wbar{Q}q} 
	        - ({\delta}B_{\wbar{Q}Q} + {\delta}B_{\wbar{q}q})}
		   {2M_{2\wbar{Q}q}} \,,
\end{equation}
where
\begin{align}
  \label{eq:deltaM}
  \delta M_{\wbar{Q}q} \equiv& M_{2\wbar{Q}q} - M_{1\wbar{Q}q}
\end{align}
is the difference between the kinetic ($M_2$) and rest ($M_1$) masses.
By construction, $I$ vanishes in the continuum limit, and it should be
closer to 0 for more improved actions.

The meson masses $M$ can be written as a sum of the perturbative quark
masses $m_1$ or $m_2$ and the binding energy $B$ as follows:
\begin{align}
  \label{eq:M1M2}
  M_{1\wbar{Q}q} = m_{1\wbar{Q}} + m_{1q} + B_{1\wbar{Q}q} \,,
  \qquad \qquad
  M_{2\wbar{Q}q} = m_{2\wbar{Q}} + m_{2q} + B_{2\wbar{Q}q} \,.
\end{align}
These formulas define $B_1$ and $B_2$.
Then, substituting them into Eq.~\eqref{eq:iparam}, the quark masses
cancel out, and the inconsistency parameter becomes a relation
among the binding energies
%
\begin{align}
  \label{eq:deltaB}
  \delta B_{\wbar{Q}q} =& B_{2\wbar{Q}q} - B_{1\wbar{Q}q} \,.
\end{align}
The corresponding quantities for Eqs.~\eqref{eq:deltaM},
\eqref{eq:M1M2}, and \eqref{eq:deltaB} for heavy ($\wbar{Q}Q$) and
light ($\wbar{q}q$) quarkonium are defined similarly.
Because light quarks always have $ma\ll1$, the $\mathrm{O}((ma)^2)$ distinction between rest
and kinetic mass is negligible.
We therefore omit $\delta M_{\bar{q}q}$ (or $\delta B_{\bar{q}q}$) when forming~$I$.
\begin{figure}[t!]
  \centering
  \vspace*{-5mm}
  \includegraphics[width=0.7\textwidth]{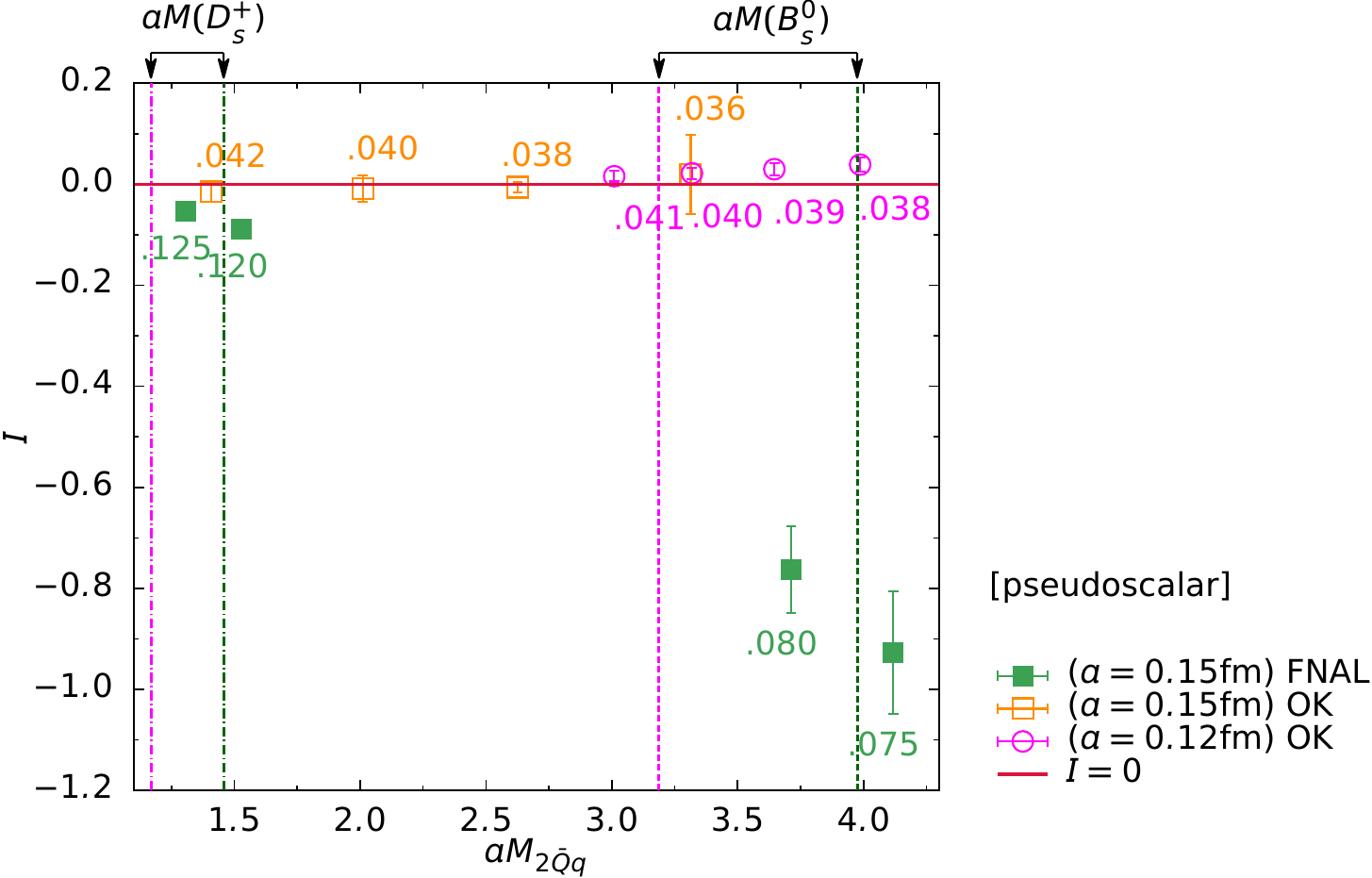}
  \caption{ Inconsistency Parameter $I$. Data labels denote $\kappa$
    values.  The square (green and orange) represents old data sets at
    $a=0.15\;$fm and circle (magenta) represents new data sets at
    $a=0.12\;$fm.  Vertical lines represent physical masses of $B^0_s$
    (dotted) and $D_s^+$ (dash-dotted) mesons. Near the $B^0_s$ meson
    mass, $I$ almost vanishes for the OK action, but for the Fermilab
    action it does not. This behavior suggests the OK action is
    significantly closer to the continuum limit.}
  \label{fig:iparam}
\end{figure}

Considering the non-relativistic limit of quark and antiquark system, for
$S$-wave case, the spin-independent binding-energy difference can be expressed as 
follows~\cite{Kronfeld,Bernard}:
\begin{align}
  \label{eq:NRdeltaB}
  \delta B_{\wbar{Q}q} 
    &= \frac{5}{3} \frac{\expv{\bm{p}^2}}{2\mu_2} \Big[ \mu_2 
       \Big(\frac{m_{2\wbar{Q}}^2}{m_{4\wbar{Q}}^3} 
          + \frac{m_{2q}^2}{m_{4q}^3} \Big) -1 \Big]
      + \frac{4}{3} a^3 \frac{\expv{\bm{p}^2}}{2\mu_2} 
        \mu_2 \left[w_{4\wbar{Q}} m_{2\wbar{Q}}^2 + w_{4q} m_{2q}^2 \right]
        + \mathrm{O}(\bm{p}^4) \,,
\end{align}
where $\mu_2^{-1} = m_{2\wbar{Q}}^{-1} + m_{2q}^{-1}$, and $m_2$, $m_4$,
and $w_4$ are defined through the quark dispersion relation analog of
Eq.~\eqref{eq:disp}.
Equation~\eqref{eq:NRdeltaB} holds for the quarkonium $\delta
B_{\wbar{Q}Q}$ too.
The leading contribution of $\mathrm{O}(\bm{p}^2)$ in $\delta{B}$
vanishes when $m_4=m_2$ and $w_4=0$, also for orbital angular momenta beyond the $S$~wave~\cite{Bernard}.
The OK action matches $m_4=m_2$ and $w_4=0$, so the two expressions in
square brackets vanish (at the tree level), leaving $I \sim \bm{p}^4 \approx 0$.

The result for the pseudoscalar channel is shown in
Fig.~\ref{fig:iparam}.
We find that $I$ is close to 0 for the OK action even in the mass
region where the Fermilab action produces very large $|I| \approx 1$.
This outcome provides good numerical evidence for the improvement expected
with the OK action.
It also shows that the new data set with the coarse ($a =
0.12\;$fm) ensemble data covers the $B_s^{0}$~mass.

\section{Hyperfine Splittings}
\begin{figure}[t]
  \centering
  \subfloat[Quarkonium]{
    \includegraphics[width=.45\textwidth]{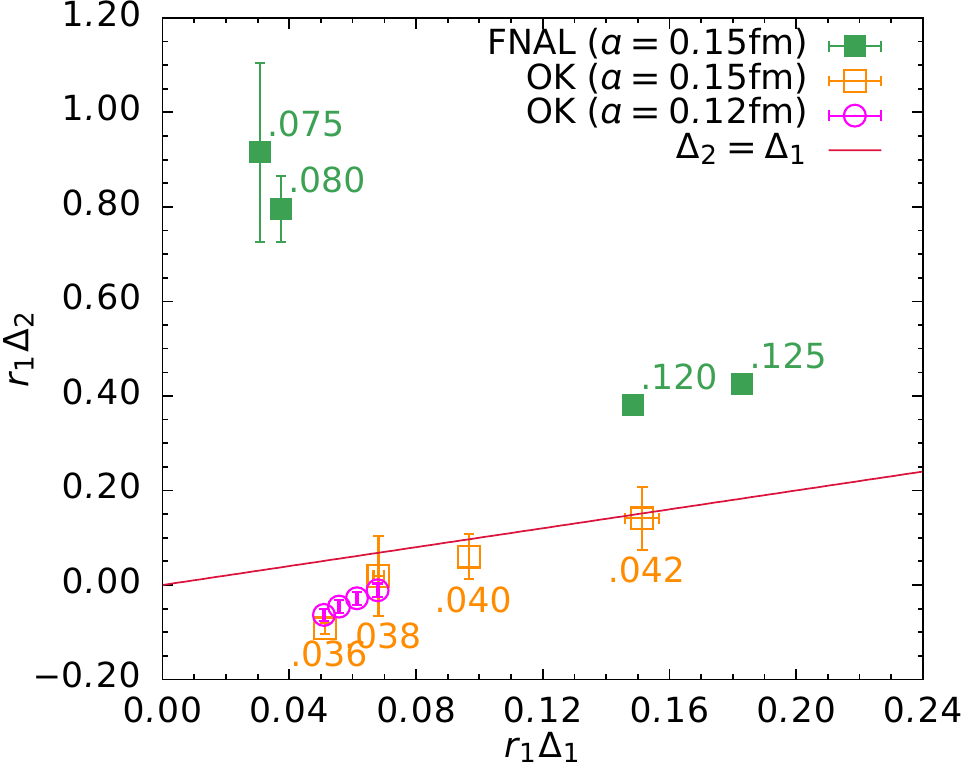}}
  \hfill
  \subfloat[Heavy-light meson]{
    \includegraphics[width=.45\textwidth]{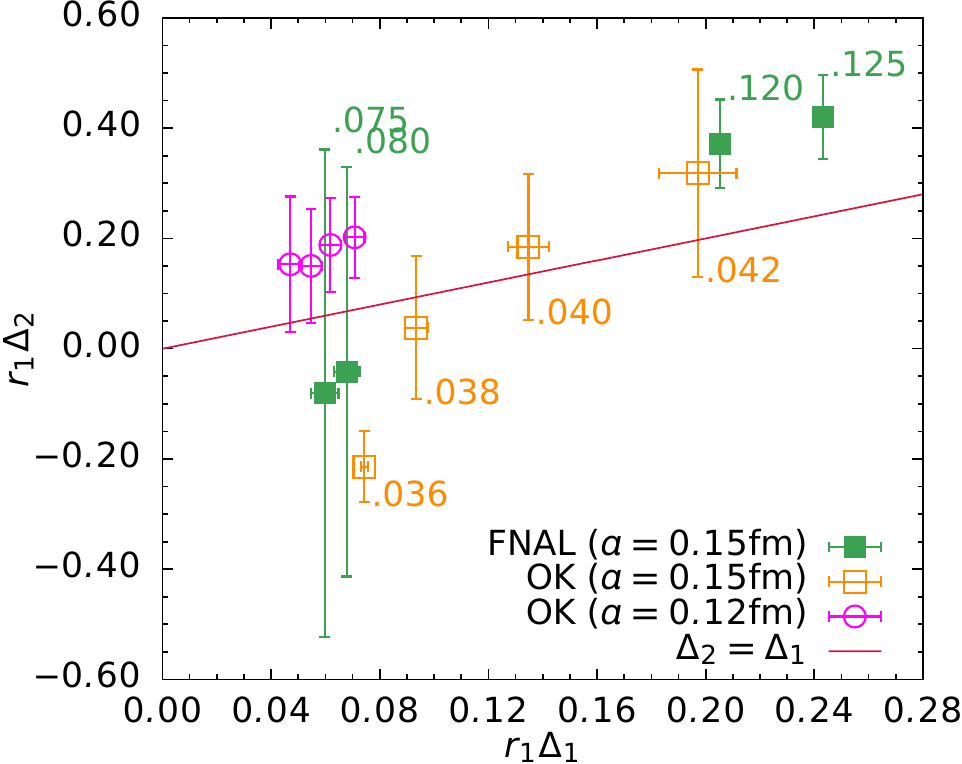}}
  \caption{Hyperfine splitting obtained from the kinetic mass vs.\ that obtained from the rest mass.}
  \label{fig:hfs}
\end{figure}
The hyperfine splitting $\Delta$ is defined to be the difference in mass
between the vector ($M^\ast$) and pseudoscalar ($M$) mesons:
\begin{equation}
  \Delta_1 = M_1^{\ast} - M_1 \,,
  \qquad
  \Delta_2 = M_2^{\ast} - M_2 \,.
\end{equation}
The hyperfine splitting of the kinetic mass ($\Delta_2$) has a larger
error than that of the rest mass ($\Delta_1$), mainly because the
kinetic mass requires correlators with $\bm{p}\neq\bm{0}$, which are noisier
than $\bm{p}=\bm{0}$.
Interestingly, with the OK action the statistical error is about $1/6$ of that with the Fermilab action,
as one can see in Fig.~\ref{fig:hfs}.
Thus, the OK action is not only more accurate in the sense of improved action but also statistically more 
precise.
%
%
%
%
%
From Eq.~\eqref{eq:deltaB}, we have
\begin{align}
  \Delta_2 &= \Delta_1 + \delta{B^{\ast}} - \delta{B} \,.
\end{align}
Spin-independent contributions to the binding energies cancel, 
so the difference in hyperfine splittings $\Delta_2 - \Delta_1$
diagnoses the improvement of spin-dependent $\mathrm{O}(\bm{p}^4)$
terms.
As one can see in Fig.~\ref{fig:hfs}, the OK action shows clear
improvement for quarkonium: the OK results lie close to the continuum limit
$\Delta_2 = \Delta_1$ (the red line).
The heavy-light results do not deviate much from the line
$\Delta_2=\Delta_1$ even with the clover action, and remain in good shape
with the OK action.

\section{Conclusion and Outlook}
The results for the inconsistency parameter show that the OK action
improves the $\mathrm{O}(\bm{p}^4)$ effects, in practice as well as in theory.
The hyperfine splitting shows that the OK action significantly
improves the higher-dimension chromomagnetic effects on the quarkonium
spectrum.
For the heavy-light system, the data for the hyperfine splittings at
$0.15\;$fm suffer from statistical errors that are too large to draw
any firm conclusion.

The SWME Collaboration plans to determine $|V_{cb}|$ by calculating
$B\to D^{(*)}\ell\nu$ semi-leptonic form factors with the OK action
and commensurately improved currents.
For this purpose, a project to obtain the improved current relevant to
the decay $B\to D^*\ell\nu$ at zero recoil is
underway~\cite{JAB:LAT2014}.
Another component of this plan is to calculate the 1-loop coefficients
for $c_B$ and $c_E$ in the OK action.
A highly optimized conjugate gradient inverter using QUDA is under
development~\cite{JANG:LAT2013}.

\section{Acknowledgments}

C.D. is supported in part by the U.S.\ Department of Energy under
grant No.\ DE-FC02-12ER-41879 and the U.S.\ National Science
Foundation under grant PHY10-034278.
A.S.K. is supported in part by the German Excellence Initiative and
the European Union Seventh Framework Programme under grant agreement
No.~291763 as well as the European Union's Marie Curie COFUND program.
Fermilab is operated by Fermi Research Alliance, LLC, under Contract
No.\ DE-AC02-07CH11359 with the United States Department of Energy.
The research of W.~Lee is supported by the Creative Research
Initiatives Program (No.~2014001852) of the NRF grant funded by the
Korean government (MEST).  W.~Lee would like to acknowledge the
support from KISTI supercomputing center through the strategic support
program for the supercomputing application research
[No.~KSC-2013-G2-005].
Part of computations were carried out on the DAVID GPU clusters at
Seoul National University.  J.A.B. is supported by the Basic Science
Research Program of the National Research Foundation of Korea (NRF)
funded by the Ministry of Education (2014027937).
%

\bibliography{refs}

\end{document}

%% file: tables/ensemble.tex
\begin{table}[b!]
\tabcolsep 3pt
\centering
\renewcommand{\arraystretch}{1.2}
\begin{tabular}{ c | c c c c c c c c }
  \hline \hline
  $a (\text{fm})$ & $N_L^3 \times N_T$
  & $\beta$
  & $a m_l$
  & $a m_s$
  & $u_0$
  & $a^{-1} (\text{GeV})$
  & $N_\text{conf}$
  & $N_{t_\text{src}}$\\
  \hline
    0.12
  & $20^{3} \times 64$
  & 6.79
  & 0.02
  & 0.05
  & 0.8688
  & 1.683$^{+43}_{-16}$
  & 484
  & 6 \\
    0.15
  & $16^{3} \times 48$
  & 6.60
  & 0.029
  & 0.0484
  & 0.8614
  & 1.350$^{+35}_{-13}$
  & 500
  & 4 \\
  \hline \hline
\end{tabular}
\caption{Parameters of the MILC asqtad ensembles with $N_f=2+1$ flavors~\cite{Bernard}. }
  \label{tbl:ensemble}
\end{table}

%% file: main.bbl
\providecommand{\href}[2]{#2}\begingroup\raggedright\begin{thebibliography}{10}

\bibitem{Wlee}
J.~A. Bailey, Y.-C. Jang, and W.~Lee {\em PoS} {\bf LATTICE2014} (2014) 371.

\bibitem{JWL}
J.~A. Bailey {\em et~al.} {\em Phys. Rev.} {\bf D89} (2014) 114504,
  [\href{http://xxx.lanl.gov/abs/1403.0635}{{\tt 1403.0635}}].

\bibitem{GS}
P.~Gambino and C.~Schwanda {\em Phys. Rev.} {\bf D89} (2014) 014022,
  [\href{http://xxx.lanl.gov/abs/1307.4551}{{\tt 1307.4551}}].

\bibitem{OK}
M.~B. Oktay and A.~S. Kronfeld {\em Phys. Rev.} {\bf D78} (2008) 014504,
  [\href{http://xxx.lanl.gov/abs/0803.0523}{{\tt 0803.0523}}].

\bibitem{EKM}
A.~X. El-Khadra, A.~S. Kronfeld, and P.~B. Mackenzie {\em Phys. Rev.} {\bf D55}
  (1997) 3933--3957, [\href{http://xxx.lanl.gov/abs/hep-lat/9604004}{{\tt
  hep-lat/9604004}}].

\bibitem{Follana:2006rc}
E.~Follana {\em et~al.} {\em Phys. Rev.} {\bf D75} (2007) 054502,
  [\href{http://xxx.lanl.gov/abs/hep-lat/0610092}{{\tt hep-lat/0610092}}].

\bibitem{Collins}
S.~Collins, R.~Edwards, U.~M. Heller, and J.~Sloan {\em Nucl. Phys. B Proc.
  Suppl.} {\bf 47} (1996) 455--458,
  [\href{http://xxx.lanl.gov/abs/hep-lat/9512026}{{\tt hep-lat/9512026}}].

\bibitem{Kronfeld}
A.~S. Kronfeld {\em Nucl. Phys. B Proc. Suppl.} {\bf 53} (1997) 401--404,
  [\href{http://xxx.lanl.gov/abs/hep-lat/9608139}{{\tt hep-lat/9608139}}].

\bibitem{MBO:LAT2010}
C.~DeTar, A.~Kronfeld, and M.~Oktay {\em PoS} {\bf LATTICE2010} (2010) 234,
  [\href{http://xxx.lanl.gov/abs/1011.5189}{{\tt 1011.5189}}].

\bibitem{Bernard}
C.~Bernard {\em et~al.} {\em Phys. Rev.} {\bf D83} (2011) 034503,
  [\href{http://xxx.lanl.gov/abs/1003.1937}{{\tt 1003.1937}}].

\bibitem{Bazavov:RevModPhys.82.1349}
A.~Bazavov {\em et~al.} {\em Rev. Mod. Phys.} {\bf 82} (2010) 1349--1417,
  [\href{http://xxx.lanl.gov/abs/0903.3598}{{\tt 0903.3598}}].

\bibitem{Wingate}
M.~Wingate, J.~Shigemitsu, C.~T. Davies, G.~P. Lepage, and H.~D. Trottier {\em
  Phys. Rev.} {\bf D67} (2003) 054505,
  [\href{http://xxx.lanl.gov/abs/hep-lat/0211014}{{\tt hep-lat/0211014}}].

\bibitem{JAB:LAT2014}
J.~A. Bailey, Y.-C. Jang, W.~Lee, and J.~Leem {\em PoS} {\bf LATTICE2014}
  (2014) 389.

\bibitem{JANG:LAT2013}
Y.-C. Jang, J.~A. Bailey, W.~Lee, C.~DeTar, A.~S. Kronfeld, and M.~B. Oktay
  {\em PoS} {\bf LATTICE2013} (2014) 030,
  [\href{http://xxx.lanl.gov/abs/1311.5029}{{\tt 1311.5029}}].

\end{thebibliography}\endgroup
